\documentclass[doublecol,figures]{epl2}
\usepackage{amssymb,epsfig,setspace,graphicx}  
\usepackage{amsmath}

\title{Comment on ``New analytic solution of Schr\"{o}dinger's equation"}

\author{Alexander Moroz}
\institute{Wave-scattering.com}

\pacs{03.65.-w}{Quantum mechanics}

\begin{document}

\maketitle

A great deal of our understanding of tunneling and many other important physics phenomena
and devices comes from the standard JWKB method. Therefore, any improvement of the JWKB method, 
such as that reported by Eleuch et al \cite{ERM}, may have far reaching applications. 
It is well known that formal substitution of the wave function $\Psi=e^{\phi(x)}$
with $\phi(x)= (i/\hbar) \int_{x_0}^x f(y)\, dy$ into the Schr\"{o}dinger equation (SE)
with a potential $U$,
\begin{equation}
\Psi''+\frac{2m}{\hbar^2}\, (E-U)\Psi=0,
\label{scheq}
\end{equation}
yields the Riccati equation \cite{Lng,MiG}
\begin{equation}
-i\hbar f'(x)+f^2(x)=p^2(x). 
\label{sc4}
\end{equation}
Here $p(x) =\sqrt{2m(E-U)}$ is the classical momentum, $m$ is mass, $E$ is energy, 
and $\hbar$ is the Planck constant. 
If one substitutes $f(x) =f_0(x)+\eta(x)$ into eq. (\ref{sc4}), assuming $\eta(x)$ to be a 
correction, the leading order satisfies the usual JWKB condition $f_0^2(x)=~p^2$. 
Hence the first-order contribution
to $\phi(x)$ is $\phi_0(x)=(i/\hbar) \int_{x_0}^x p(y)\, dy = (i/\hbar) S$ 
involving the classical action $S$. For the correction $\eta(x)$ one arrives at
\begin{equation}
-i\hbar\eta'(x)- i\hbar p'(x)+ 2p(x)\eta(x)+ \eta^2(x) = 0.
\label{sc6}
\end{equation}
On solving the equation {\em iteratively} by expansion into powers of $\hbar$, 
the next-to-leading order contribution $\phi_1= \ln (p/\hbar)^{-1/2}$ leads to 
the standard JWKB approximation,
\begin{equation}
\Psi_{\scriptscriptstyle\rm JWKB}(x) \sim k^{-1/2}(x) \exp \left(i \int_{x_0}^x k(x)\, dx \right),
\label{wp}
\end{equation}
where $k=p/\hbar$ is the wave vector. The other linearly independent solution 
is obtained by choosing the wave vector $k$ with a {\em minus} sign.

The novelty of the approach by Eleuch et al \cite{ERM} is to determine
the next-to-leading order contribution differently: not iteratively
but as a closed analytic solution 
of a {\em linearized} equation (\ref{sc6}) without the quadratic term $\eta^2(x)$.
Remarkably, such a closed analytic solution comprises the full 
next-to-leading contribution of the JWKB method
and, at least partially, the contributions of {\em all} higher orders 
were eq. (\ref{sc6}) solved iteratively \cite{ERM}.
The closed analytic solution is thus a partial sum of an infinite
asymptotic series.

The linearized equation was written in two alternative forms 
[{\em cf.} eqs. (17) and (32) of \cite{ERM} in a slightly different notation]
\begin{eqnarray}
-i\hbar\eta'(x)-i \hbar p'(x)+ 2p(x)\eta(x) = 
\nonumber\\
-i\hbar f'(x)+ 2p(x)[f(x)-p(x)] = 0.
\label{iwkb}
\end{eqnarray}
Being first-order ordinary differential equations, their general solution
contains an {\em arbitrary} integration constant. The most general solution to the first 
of eqs. (\ref{iwkb}) is 
\begin{equation}
\eta(x)= -\int_{x_0}^x [p'(y)+C_1]\, 
 e^{\frac{2i}{\hbar} [S(y)-S(x)]}\, dy,
\label{etagsl}
\end{equation}
where $C_1$ is such an integration constant. 
Analogously, the most general solution to the second of eqs. (\ref{iwkb}) is
\begin{equation}
f(x)= \frac{2i}{\hbar}\, e^{-\frac{2i}{\hbar} S(x)}
          \left(C_2+ \int_{x_0}^x p^2(y)\, e^{ \frac{2i}{\hbar} S(y)}\,  dy\right).
\label{etaslfaf}
\end{equation}
One assumes here $S(x_0)=0$, or that the lower limit in the integral
defining the action $S$ is $x_0$.
The respective constants can be fixed by the requirement
that $\Psi(x) \sim \exp (\tfrac{i}{\hbar} \, S)$ in the region with $p'\equiv 0$. The latter
requires $C_1=0$ in eq. (\ref{etagsl}) and $C_2=\tfrac{\hbar}{2i}\, p(x_0)$ in eq. (\ref{etaslfaf}).

Our {\em first} observation is that eqs. (8) and (34) of ref. \cite{ERM} are equivalent,
which can be demonstrated by integration by parts, only under 
the above choice of integration constants. The integration constants were omitted 
in refs. \cite{ERM,ER} and the solution (\ref{etaslfaf}) 
with $C_2=0$ was suggested to be applied to the case of a {\em step} potential.
However, the integral in (\ref{etaslfaf}) contributes to $f(x)$ an {\em unphysical}
term $- p\exp [-2ik(x-x_0)]$ in any $p'= 0$ region.
The unphysical term cancels out only with the correct choice of $C_2$.
Furthermore, an exponentiation of the integral of the rhs of eq. (35) of \cite{ERM} 
cannot yield the final result of eq. (36).

Our {\em second} observation is that the improved JWKB method {\em cannot} 
be demonstrated in the case of a step potential. 
In any region with a {\em constant} momentum, and hence $p'(x)\equiv 0$, 
the two linearly independent {\em exact} solutions $\exp (\pm ik_jr)$
of the SE (\ref{scheq}) are {\em fully} reproduced by both 
the standard and improved JWKB approximations.
In the latter case this is exemplified in that the correction $\eta(x)$ given 
by eq. (\ref{etagsl}) with $C_1=0$ vanishes 
in the regions where $p'\equiv 0$. Hence an improved JWKB {\em cannot} provide anything 
new for a step potential or a {\em rectangular potential barrier}, 
because then $p'=0$, and hence $\eta=0$, everywhere, except for the step.
Obviously, by performing a proper analysis of the {\em step} potential by
requiring the continuity of logarithmic derivatives of $\Psi(x)$ given by eq. (37)
of ref. \cite{ERM} across the barrier step there {\em cannot} be any
difference between, on one hand, solving the SE exactly 
and, on the other hand, by making use
of either standard or improved JWKB method (because $\eta(x)\equiv 0$ before and after the step).
Consequently, the results plotted in fig. 4 of ref. \cite{ERM} are a mere artifact of 
an improper treatment of the step potential. Eqs. (16) of a follow up \cite{ER} of 
two of the authors suggest that a counter-propagating reflected plane wave was 
disregarded on both sides of the step. 

Analytically, the above conclusions are supported by looking
at the equations which {\em exact} solutions are given SE approximations \cite{Lng}. 
$\Psi_{\scriptscriptstyle\rm JWKB}$ of the JWKB method satisfies
\begin{equation}
\Psi''(x)+ [k^2(x) + W(x)]\, \Psi(x)=0,
\label{mjwkbe}
\end{equation}
with an ``{\em error}" function $W$ [{\em cf.} the SE (\ref{scheq})] given by 
\begin{equation}
W_{\scriptscriptstyle\rm JWKB}(x) =\frac{3[k'(x)]^2}{4k^2(x)} - \frac{k''(x)}{2k (x)}
= -\frac{1}{2}\,\{S,x\} 
=\frac{T''}{T},
\nonumber
\end{equation}
where $\{S,x\}$ is the {\em Schwarzian derivative} of $S(x)$ and
$T= S'^{-1/2} = k^{-1/2}(x)$ \cite{Lng}. 
Because $W_{\scriptscriptstyle\rm JWKB}$ {\em diverges} for $k\rightarrow 0$, 
each turning point is a {\em singular} point 
of the differential equation (\ref{mjwkbe}). This is the very reason why the JWKB solutions 
$\Psi_{\scriptscriptstyle\rm JWKB}(x)$ of eq. (\ref{wp}) are branches of a
{\em multivalued} function in the proximity of a turning point and
the so-called {\em connection formulas} are required to determine a 
solution on both sides of the turning point. There are the turning points 
where one should look for improvements over the JWKB method.

The approximation of Eleuch et al \cite{ERM} is found to be {\em exact}
solution of eq. (\ref{mjwkbe}) with $W=\eta^2/\hbar^2$. 
Obviously, $W\equiv 0$ in any $p'=0$ region ({\em cf.} the step potential).
In general eq. (\ref{etagsl}) implies that $|\eta(x)|\le \int_{x_0}^x |p'(y)|\, dy$.
Assuming a finite integration range and not pathologically oscillating $p'(x)$,
$\eta(x)$ should be {\em bounded} for any bounded $p(x)$.
This brings about a significant improvement over the JWKB method,
because turning points now become {\em regular} points of the differential equation (\ref{mjwkbe}).
Consequently, there is no longer a catastrophic failure 
of the improved approximation around turning points. Solutions of
eq. (\ref{mjwkbe}) yield a good approximation to the solutions of the SE (\ref{scheq}) 
whenever $|W|\ll |k^2(x)|$. Because $W$ is {\em not} guaranteed to go to zero for $k(x)\rightarrow 0$ 
for the approximation of Eleuch et al \cite{ERM}, turning points may still cause some problems.
A quantitative assessment of the deviation at turning points remains an open problem.

{\em Thirdly}, the improved JWKB method by Eleuch et al \cite{ERM} does not in general conserve 
probability for a real $p$. One has 
\begin{equation}
- \frac{i\hbar}{2m} \frac{d}{dx} \left(
\Psi^* \, \frac{d\Psi}{dx} - \frac{d\Psi^*}{dx} \Psi \right) 
        = \frac{i (\eta^2-\eta^{*2})}{2m\hbar}\,|\Psi|^2,
\nonumber
\end{equation}
which is valid also for imaginary $p$. In arriving at the result
we have used that $f'= - \tfrac{2i}{\hbar}\, p\eta$ and $p^2=p^{*2}$. 
Hence the probability is conserved if and only if $\eta^2=\eta^{*2}$
({\em e.g.} in the region where $p'\equiv 0$).
The conservation is obviously {\em violated} in the generic case when $\eta\ne 0$ is 
neither real nor purely imaginary. Although the violation should be 
kept in mind when applying the approximation to real problems,
it need not to be necessarily a serious issue.
This is because the improved JWKB approximation comprises the first two iteration orders 
yielding $\Psi_{\scriptscriptstyle\rm JWKB}$ of the JWKB method.
However $\Psi_{\scriptscriptstyle\rm JWKB}$ conserves probability if 
the action $S$ is {\em real}. 
Therefore, any violation of the probability conservation can affect only higher orders.

To put the work by Eleuch et al \cite{ERM} into perspective, 
the method of {\em comparison} equations yields approximations which 
(i) do conserve probability and (ii) the error function $W$ in eq. (\ref{mjwkbe}) 
goes to zero for $p\rightarrow 0$ \cite{Lng,MiG}.
However, the latter are specially designed for the crossing of turning points.
They involve typically {\em transcendental} ({\em e.g.} Bessel and parabolic cylinder) functions 
of argument $S/\hbar$ and of the order determined by the order of a turning point \cite{Lng,MiG}.
They reduce to $\Psi_{\scriptscriptstyle\rm JWKB}(x)$
in the regions where $p'\equiv 0$ only for $|S/\hbar|\rightarrow \infty$. Hence the simplicity of performing various 
integrals with an exponential function as in the standard and improved JWKB methods is lost. 
Therefore, the approximation by Eleuch et al \cite{ERM} may be a useful compromise
between $\Psi_{\scriptscriptstyle\rm JWKB}(x)$ and the method of 
{\em comparison} equations \cite{Lng,MiG}. It remains to be seen
if it justifies the above expectations. The preliminary results for the energies
above different barriers are promissing \cite{ERM}.

\acknowledgments

Continuous support of MAKM is largely acknowledged.


\end{document}